# Stress-dependent electrical transport and its universal scaling in granular materials

Chongpu Zhai[1,2] [*], Yixiang Gan[1], Dorian Hanaor[3], Gwénaëlle Proust[1]

[1]School of Civil Engineering, The University of Sydney, Sydney, Australia
[2]Hopkins Extreme Materials Institute, The Johns Hopkins University, United States
[3]Fachgebiet Keramische Werkstoffe, Technische Universität Berlin, Germany

E-mail: czhai3@jhu.edu

**Abstract**

We experimentally and numerically examine stress-dependent electrical transport in granular materials to elucidate the origins of their universal electrical response. The dielectric responses of granular systems under varied compressive loadings consistently exhibit a transition from a resistive plateau at low frequencies to a state of nearly constant loss at high frequencies. By using characteristic frequencies corresponding to the onset of conductance dispersion and measured direct-current resistance as scaling parameters to normalize the measured impedance, results of the spectra under different stress states collapse onto a single master curve, revealing well-defined stress-independent universality. In order to model this electrical transport, a contact network is constructed on the basis of prescribed packing structures, which is then used to establish a resistor-capacitor network by considering interactions between individual particles. In this model the frequency-dependent network response meaningfully reproduces the experimentally observed master curve exhibited by granular materials under various normal stress levels indicating this universal scaling behaviour is found to be governed by i) interfacial properties between grains and ii) the network configuration. The findings suggest the necessity of considering contact morphologies and packing structures in modelling electrical responses using network-based approaches.

**Keywords**: RC network, contact mechanics, granular materials, electrical transport, universal scaling.

# 1. Introduction

The power-law scaling of dielectric properties with frequency, known as the 'universal dielectric response' (UDR) [1], has been observed in a wide range of materials, including disordered ceramics [2, 3], ion/electron-conducting glasses [3-5], amorphous semiconductors [4, 6, 7], metal powders [8] and nanoparticles [9, 10]. This frequency-dependent behaviour is of importance in diverse applications including material characterization [2, 4], battery optimisation [11, 12], and electronic sensors [13, 14]. Investigations of scaling in AC (alternating current) conduction have been conducted with simple equivalent circuit containing paralleled electrical components [8, 15], square random resistor-capacitor (RC) networks [8, 16-18], random-barrier models [4, 19, 20] and effective-medium approximation [20-22]. For different types of ionic [4, 23] and electronic [10] conduction, the dependence of conductivity ($\sigma$) on frequency ($\omega$) under a wide range of temperature conditions demonstrates a single master curve, suggesting the validity of the time-temperature superposition principle (TTS) described by the scaling law

$$\sigma(\omega)/\sigma^* = f(\omega/\omega^*), \quad (1)$$

where $f$ is a temperature-independent scaling function, $\sigma^*$ the dc (direct current) conductivity and $\omega^*$ the characteristic angular frequency corresponding to the onset of conductivity dispersion. This scaling behaviour may stem from various transport mechanisms in terms of timescales [4, 5, 19]. At different temperatures, $T$, all characteristic times, $\tau^* = 1/\omega^* \propto \exp(T^{-1})$, exhibit the same rate of ion hopping associated with activation energy, indicating the scaling of dynamic processes is temperature-independent [6, 19, 24, 25]. Additionally, the proportionality of $\sigma^* \propto \omega^* \propto 1/\tau^*$ can be derived through the Barton-Nakajima-Namikawa relation, revealing TTS behaviour [26]. Such TTS behaviour has also been observed for granular nanocomposites [9, 10, 27] and granular metal films [7, 28], where thermally-activated electron tunnelling among nanoparticles dominate electronic conduction, with $\tau^* \propto \sigma^* \propto \exp(T^{-1/2})$. In both ionic and electronic conduction $\tau^*$ and $\sigma^*$ exhibit Arrhenius-type dependence on $T^n$ [4, 25], giving rise to the inherent spectral features with shape of the impedance spectra unaffected by temperature. The corresponding master curves for various materials are found to exhibit power law regions with exponents ranging from 0.6 to 1 within a certain frequency range [1, 4, 5, 10, 19, 29, 30]. At high frequencies, a regime of nearly constant loss (NCL) with the dispersion exponent approaching unity can be observed [19, 31, 32].

Scaling behaviour described by TTS can be expected with the fulfilment of $\tau^* \propto \sigma^* \propto \exp(\delta^n)$, where $\delta$ denotes the mechanical loading. Stress-dependent scaling in granular materials has seldom been studied and the mechanisms involved are not well understood. Moreover, few studies have focused on the underlying reasons for shape deviation in the TTS master curve [4, 10, 30]. In this letter, we examine the frequency-dependent electrical responses of randomly packed granular materials under various compressive loads. By considering inter-granular electrical transport within networks established from prescribed packing structures, we elucidate the fundamental drives shaping the master curve for various topological systems.

## 2. Experimental observations

Monosized stainless steel spheres (AISI 304, with precision grade G200) were randomly packed in a nonconductive cylinder ($Al_2O_3$) of 85 mm in height, $H$, 98mm in diameter, $D$, with circular steel plates as top and bottom electrodes. Impedance spectra were measured using an impedance analyser (Agilent 4294A) from 40 Hz to 10 MHz, and plotted as a function of frequency $\omega$ as shown in Fig. 1. We performed isobaric measurements for spheres of three different diameters ($d$), 1 mm (with applied pressures up to 434.43 kPa), 5 mm and 10 mm (1.91 kPa to 33.71 kPa). Within the applied pressure range, all obtained impedance values are considerably larger than the total resistance (< 10 mΩ) of the measurement system excluding the packed bed under testing. Here, we consider particles of three different sizes in order to vary the resulting inter-particle force for a given applied mechanical compression and also to achieve various roughness-to-diameter ratios. To minimize current-induced local welding, we applied an AC signal with a peak-to-peak value of 200 mV, for which linear ohmic behaviour can be observed in current-voltage responses for the considered loading range [8, 33, 34], this is detailed in Supplementary Materials. Controlled measurements were performed to exclude parasitic resonance signals from cabling and connections, using measured Nyquist plots [35].

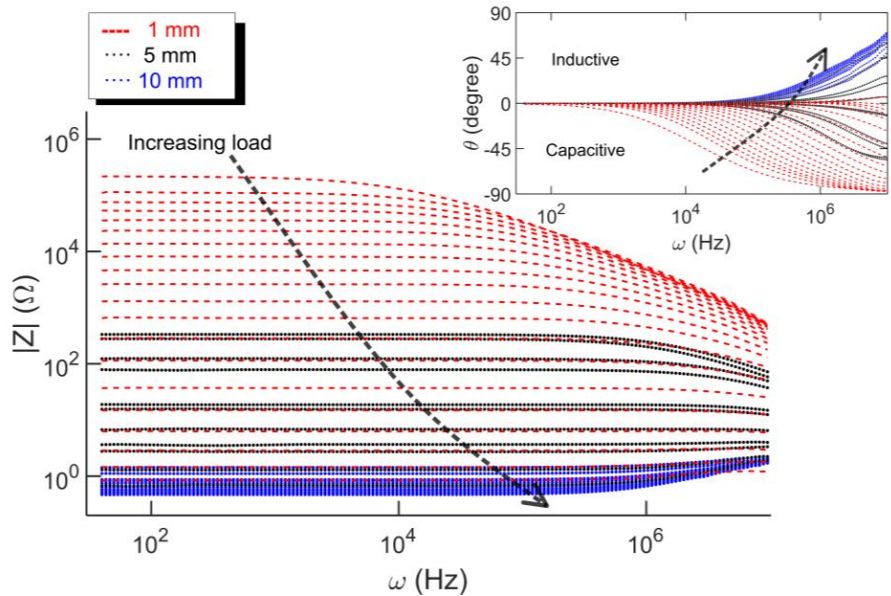

**FIG. 1.** Typical measured impedance modulus, $|Z|$, as a function of frequency, $\omega$. Insert shows the impedance phase, $\theta$, versus frequency. Results for spheres of three different sizes are presented using red, black and blue lines, corresponding to, 1 mm, 5 mm and 10 mm diameter spheres, respectively.

The measured impedance is characterised by a transition above a critical (angular) frequency $\omega_c$ from a low-frequency dc plateau to a dispersive region. At high frequencies (above 0.01 MHz for 1 mm grains and above 0.3 MHz for 5 mm and 10 mm grains), the impedance angle is found to rise with the applied mechanical load. Specifically, decreasing and increasing impedance modules were observed for 1 mm and 10 mm grains, respectively. For 5 mm grains, the impedance angle increases from negative to positive values indicating a capacitive-to-inductive phase transition. In order to illuminate the observed phase transition, the studied systems are simplified as networks formed by a large number of electrical components [3, 16, 17, 36]. Here, an electrical network is established by incorporating stress-dependent electrical interactions between spheres. Adjacent particles can be treated as a capacitor. The interaction between two contacting particles will evolve from a capacitor ($X_T$, due to the existing oxide layer [37]), to an effective resistor ($R_T$) in parallel with an effective capacitor ($X_T$), under increasing compression [8, 34, 38-41]. This evolution is typically governed by the surface structures of interacting spheres, exhibiting multiscale geometries [42, 43]. Here, the grain surfaces were scanned using an optical surface profilometer (NanoMap 1000WLI) to obtain three-dimensional (3D) digitised topographies, shown in Supplementary Materials.

For a single contact spot ($i$), the convergence and divergence of current give rise to a constriction resistance $R_{ci}$, where electrons are transported in different mechanisms depending on contact size, e.g. via Sharvin, Wexler, and Maxwell conduction [44-48]. Additionally, tunnelling resistance $R_{ti}$, in series with $R_{ci}$, through the oxide layer and voids contributes significantly to the resistance at an established micro-contact [27, 44]. At the contact level, the effective resistance incorporates numerous electrical micro-contacts at interacting asperities. Micro-voids at contact regions in which trapped air is present form micro-capacitors (with capacitance presented by $C_i$) in parallel [34, 38, 41]. The total electrical contact impedance $Z_T$ is formulated by

$$Z_T = X_b + R_T // X_T \approx \left[ \frac{1}{\sum_i 1/(R_{ci} + R_{ti})} \right] // (\frac{j}{\omega \sum_i C_i}), \qquad (2)$$

where the operator "//" represents parallel connection, $j$ the imaginary unit, and $X_b$ the reactance of the bulk, the value of which is typically orders of magnitude smaller than that of $R_T$ and $X_T$, for the considered frequency ($\ll$ GHz) [41] . For adjacent spheres, Eq. (2) can be simply written as $Z_T = X_T \approx j/C_{eff}$, with $X_b \to 0$, $R_T \to \infty$ and $i = 1$ accounting for the effective capacitor $C_{eff}$ by the void between adjacent grains. Noticeably, large zones of metal-to-metal contact can be achieved with sufficiently high pressures, establishing continuous metallic pathways through contacting grains, leading to inductive system behaviour. Based on the interfacial properties, the AC responses of assemblies of 1 mm spheres are investigated using an RC network containing $C$ and $R//C$ components, since only capacitive dispersion was observed within the applied loading range. At higher inter-particle forces, results showing inductive properties can be similarly realised by RCL networks containing $C$ and $R//C + L$, where the operator "+" represents series connection and $L$ represents the inductance arising from strong force chains [49].

To further analyse the stress-dependent plateau values and characteristic frequency, we first identify these quantities from the recorded spectrum. As is shown in Fig. 2, the typical reactance, $Z''$, spectra for a 1 mm bed under varied compression exhibits peaks, $Z''^*$, at characteristic frequencies, $\omega^*$, marking the beginning of conduction dispersion and relaxation process. These peaks indicate that the conduction of $C$ component approaches to that of $R$ at $\omega^*$. At higher frequencies, the conduction of $C$ component further surpasses that of $R$, associated with the downward trend shown in Fig. 1. This characteristic frequency increases with load while the corresponding reactance peak diminishes in magnitude. Importantly, all $Z''$ curves follow a single master curve using $Z''^*$ and the corresponding $\omega^*$ values as scaling factors, suggesting relaxation processes is stress-independent. At characteristic frequencies, the magnitudes of $R_T^*$ and $X_T^* = 1/(\omega^* C_T^*)$, corresponding to representative values of a single interaction between a pair of spheres in the system, are similar [10, 50], thus we have $\omega^* = 1/(C_T^* R_T^*)$.

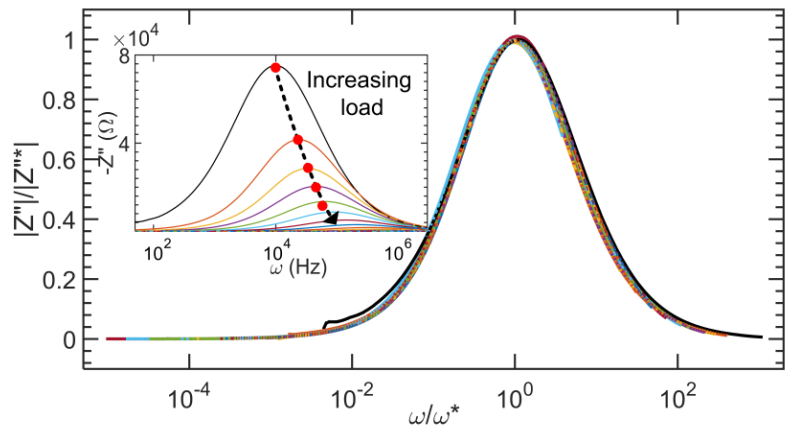

**FIG. 2.** Typical scaling of $|Z''|/|Z''^*|$ versus $\omega/\omega^*$ for 1 mm bed. The insert shows $Z''$ as a function of frequency with red dots marking $Z''^*$ peaks at $\omega^*$, for increasing loads.

The dependence of normalised plateau values, $\overline{R^*}$, on normalised applied pressure, $\overline{F^*}$, is presented in Fig. 3. The measured dc impedance of granular packing, $R^*$, is obtained by averaging over ten tests and is shown with error bars. The representative inter-particle force, $F_T^*$, and $R_T^*$, are linearly correlated to dimensionless load and resistance, respectively:

$$\overline{F^*} = \frac{4d^2(F_0 + F)}{(\pi D^4 E)}, \qquad \overline{R^*} = D^2(R - R_C)/(HdR_0), \qquad (3)$$

where $R = R^* + R_C$ is the measured resistance of the whole system, $R_C$ the total resistance of the cables and connectors, $R_0$ the theoretical resistance of the bulk material for an identical volume as that of the packed bed, $E$ the Young's Modulus of the tested material (203 GPa), $F_0$ the sum of the half weight of the granular assembly along with the top electrode, and $F$ the applied compressive force. The presented normalization in Eq. (3) was proposed based on the nature of this research problem in terms of grain size, network responses, inter-particle forces and electrical resistance. To normalise $\overline{F^*}$, the load exerted on the sphere packing, $F_0 + F$, was first expressed by dimensionless pressure, being divided by apparent area of the packing, $A = \pi D^2/4$, and the material's elastic modulus, $E$. By incorporating the ratio between the cross-sectional areas of the packing and a single sphere, $A/a = (\pi D^2/4)/(\pi d^2/4)$, the obtained apparent stress was associated with the representative inter-particle force. This ratio indicates how the apparent pressure is effectively concentrated at scattered contact spots. In a similar way, we linked the measured resistance for the entire packing to a representative inter-particle electrical resistance. The measured resistance $R - R_C$ for sphere packing was first divided by $R_0$ to obtain the relative resistance, which is a dimensionless quantity. We then considered each individual inter-particle electrical resistance, $R_i$, as a standard cuboid resistor element. With this simplification, the relative resistance of the whole packing can be effectively calculated by $(R - R_C)/R_0 = R_i(H/d)/(A/a)$, where $A/a$ denotes the number of $R_i$ elements in a single layer, which can be considered as being connected in parallel, while $H/d$ denotes the number of layers of $R_i$ in series connection.

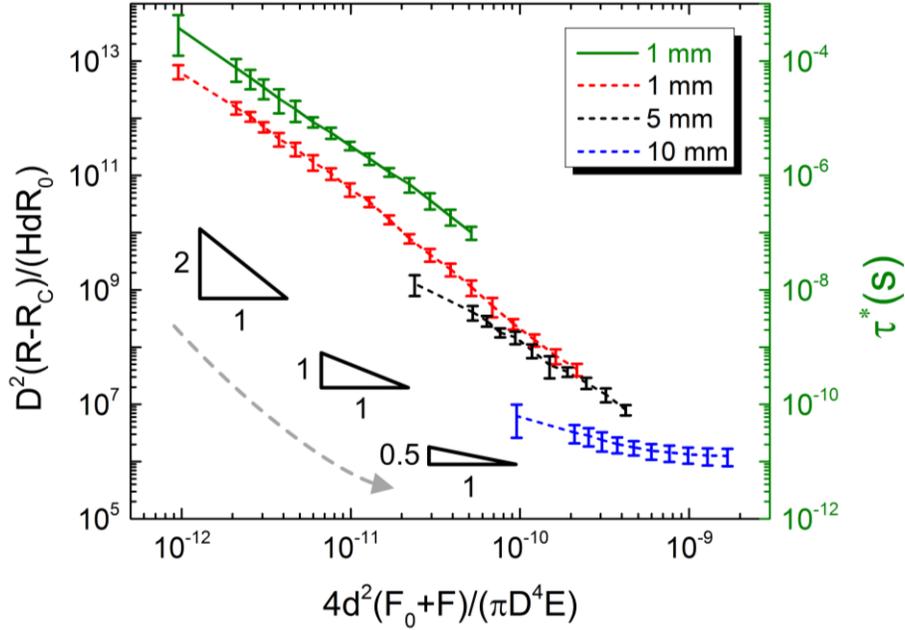

**FIG. 3.** Dependence of dimensionless $\overline{R^*}$ on $\overline{F^*}$ for assemblies of different-sized spheres shown with dashed lines with error bars. The dependence of $\tau^*$ on $\overline{F^*}$ for 1 mm diameter spheres, with pressures from 1.91 kPa to 103.36 kPa, is presented by a solid line relating to the right y-axis.

The characteristic time, $\tau^* = 1/\omega^* = C_T^* R_T^* \propto C_T^* \overline{R^*}$, is plotted as a function of $\overline{F^*}$ in Fig. 3. Straightforwardly, we can ascertain $\tau^* \propto \overline{R^*}$ due to the observed power law functions of

$$\overline{R^*} \propto \overline{F^*}^{-2.09 \pm 0.11}, \qquad \tau^* \propto \overline{F^*}^{-2.01 \pm 0.10}, \quad (4)$$

with both exponents being approximately -2. The difference between the two exponents comes from $C_T^*$, which is mainly affected by inter-particle distance [7, 28, 51] for a given particle size. A higher load results in a smaller inter-particle distance, therefore, a higher value of $C_T^*$. However, the observed close power exponent values for dependence of $\overline{R^*}$ on $\overline{F^*}$ and $\tau^*$ on $\overline{F^*}$ demonstrate that the characteristic frequency is mainly governed by $R_T^*$ for the considered experimental conditions in this work. Noted that the value of $C_T^*$ can be also affected by the surface structure of contacting spheres [34, 38, 41], material properties [19, 22, 36], grain size [8, 10] and applied load [52]. For 10 mm spheres, absolute exponent values of $0.83 \pm 0.14$ and $0.32 \pm 0.09$ can be found, corresponding to the five lowest and highest load levels, respectively. The latter exponent is comparable to that (1/3) of Hertzian contact with Holm conduction as the dominant transport mechanism. One can obtain relationships of $A_H \propto F_H^{2/3}$ and $R_H \propto A_H^{-1/2}$, respectively, based on Hertzian theory and Holm conduction, thus, having $R_H \propto F_H^{-1/3}$ [33, 45, 48], where, the subscript, $H$, indicates the Holm conduction at Hertzian contact. The experimentally observed exponent, i.e., $0.32 \pm 0.09$, agrees well with the theoretical prediction, i.e., $1/3$, for a smooth sphere-to-sphere contact. This consistency demonstrates that the electrical contact at high inter-particle forces can be described as a Hertizan contact with Holm conduction. For the considered loads in experiments, we observe a decreasing trend of the absolute exponent value, from 2 to 1/3 with increasing inter-particle force. This trend indicates the diminishing influence of surface roughness and the evolution of conduction mechanisms as discussed in previous research [33, 45, 48].

Finally, we show, in Fig. 4(b) and Fig. 5(b) with black dots, $|Z|/R^*$ versus $\omega/\omega^*$ for 1 mm grains with the load range of 1.91 kPa to 103.36 kPa, where $Z''^*$ and $\omega^*$ can be recognized for the considered frequency range. $|Z|$ curves under all loading conditions collapse onto a single master curve with $R^*$ and $\omega^*$ acting as scaling parameters, demonstrating inherent electrical features across a wide frequency range.

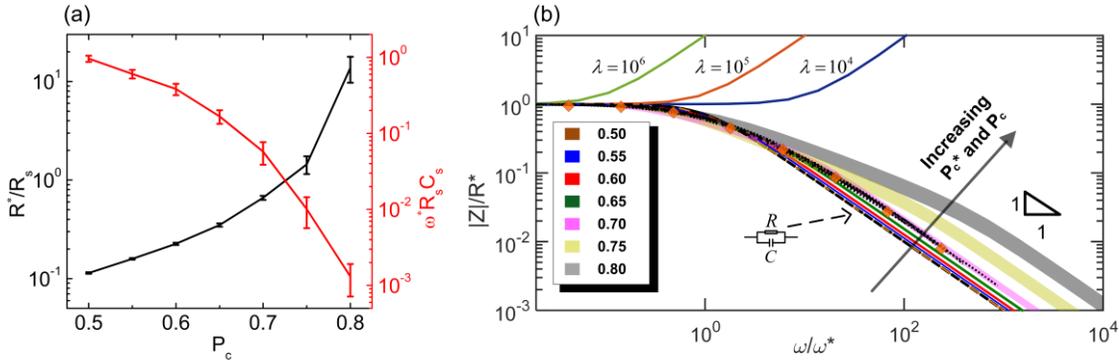

**FIG. 4.** (a) Dependence of $R^*/R_s$ and $\omega^* R_s C_s$ on $P_c$, corresponding to the left and right y-axes, respectively. (b) Scaling plots of $|Z|/R^*$ versus $\omega/\omega^*$. Experimental results for 1 mm diameter spheres are shown in black dots with the fitted master curve marked with orange diamonds. Simulation results of RC and RCL networks, respectively, with various $P_c$ and $\lambda$ are presented by shaded error bars. The AC response of simple $R//C$ equivalence is presented by the black dashed line. Each individual point shown in error bar is based on ten realizations.

### 3. Numerical simulation

To explore the origins of the intrinsic electrical characteristics observed, we incorporated the electrical interaction between spheres into prescribed random packing structures realized using a discrete element method [53]. The obtained packing is sandwiched between two conductive rigid flats, one of which is grounded and the other is raised to a potential $V(\omega)$. The impedance can be obtained by solving the complex linear equations established by applying Kirchhoff's current law to each particle, denoting a node of the network [50]. A simulation volume larger than 10×10×10 particle diameters has been recommended to reduce the size dependence of the percolation threshold [54, 55]. Here, 8000 spheres were randomly packed in a rigid container with a height-area ratio and initial volumetric packing (62%). For comparison, increasing compression in experiments resulted in the volume ratio increasing from 61.233 ± 0.859% to 62.716 ± 0.847% over ten tests. With these settings, a coordination number $N_c$ of 6.14 ± 0.11 over ten realisations is obtained, indicating the number of interacting spheres within the cut-off distance of a reference sphere [53, 56, 57]. This distance is initially set to equal the grain diameter.

For a 1 mm bed, either $C$ or $R//C$ are assigned to "interacting" pairs of spheres. Consequently, the $C$ element proportion ($P_c^*$) in a range [0, 1] results in the total capacitor proportion ($P_c$) ranging from 0.5 to 1, as $P_c = 1/(2 - P_c^*)$. With this network structure, $R//C$ type components constitute the network backbone, providing equal numbers of capacitors and resistors. Spheres in weak contact, acting effectively as capacitors, add additional capacitive phases. As a simplified assumption, all $C$ and $R$ elements are assigned identical values, $C_s$ and $R_s$, respectively, across the network without referring to the contact force distribution. This assumption is justifiable on the basis of the robustness of RC network responses with respect to microstructural details and component positions [1, 17, 19, 22, 24]. Here, the absolute value of the average contact force can be dimensionalised using Eq. (3). In order to further study the effects of component values on the resultant electrical responses, we conducted simulations with resistors and capacitors of various magnitudes, detailed in Supplementary Materials. The observed insensitivity of network responses to the properties and positions of elements is consistent with previous literature [8, 17, 22, 24].

Numerical and experimental results are compared in Fig. 4 and 5 with analyses based on a simple circuit-equivalence (a resistor and a capacitor in parallel) given for reference [8]. Higher values of $P_c$ (smaller than the percolation threshold with $P_c \approx 0.83$ [58-60]) give rise to smoother transitions from resistive plateaus to NCL regimes. The master curve obtained from experimental results is well reproduced for $P_c = 0.70$. This $P_c$ value is closely related to interfacial properties including roughness, oxide layer thickness and dominant conduction mechanism [8, 33, 42, 47]. Moreover, the resistive plateau and onset of conduction dispersion are primarily dominated by $P_c$, as shown by the dependences of $|Z|/|R_s|$ and $\omega^* R_s C_s$ on $P_c$ presented in Fig. 4(a).

To reproduce the observed inductive behaviour, RCL networks are simulated. As shown in Fig. 4(b), the onset of dispersion is found to be dominated by the dimensionless ratio, $\lambda = L/(CR^2)$, governed by interfacial properties [33, 38, 41]. Under increasing inter-particle forces, contacts evolve from $C$ to $R//C$ and then to $R//C + L$ constituting a larger proportion, e.g., 0.9 used in the presented simulations. As a result, the current will percolate through inductors, and thus, present inductive dispersion at high frequencies [47]. Note that here we simulated the inductive behaviour with different values of $L$, which resulted in distinctive electrical responses, unlike the universal scaling observed in the capacitive phase.

We have endeavoured to interpret experimental measurements with a reasonable working hypothesis based on interfacial properties. In order to investigate the role of packing structures in the universal scaling characterized by the master curve, we further applied the presented numerical framework with $P_c = 0.70$ to networks with various configurations. Rather than 2D or 3D square lattice networks [16, 19, 22, 50], we vary the cut-off distance defining the limits of particle interactions to obtain network structures with different coordination numbers [53, 55, 58]. Networks with mean $N_c$ values of $4.02 \pm 0.15$, $6.01 \pm 0.20$, $7.99 \pm 0.16$, $10.13 \pm 0.21$, and $12.08 \pm 0.16$ over ten realizations of packings were achieved, corresponding to applied cut-off distances of $0.99\,d$, $1.02\,d$, $1.11\,d$, $1.30\,d$, and $1.45\,d$, respectively. As is shown in Fig. 5, a higher $N_c$ results in a later onset of conduction dispersion, lower resistive plateau value and sharper transition from resistive to NCL regimes. The coordination number indicating the connectivity of the network structure, is associated with the effective local dimension of current pathways, which fundamentally determines the conduction properties of network structures [58, 59], and thus the transition from resistive plateau to dispersive conduction as frequency increases.

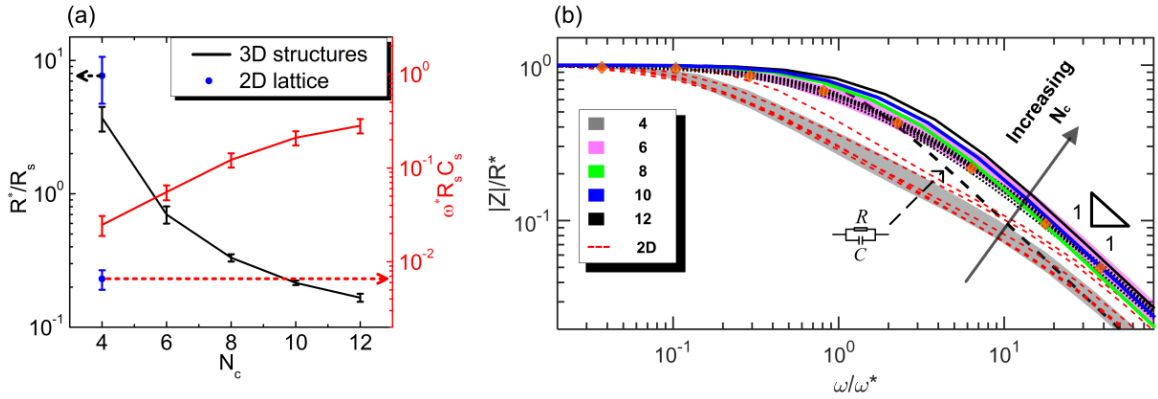

**FIG. 5.** (a) Dependence of $R^*/R_s$ and $\omega^* R_s C_s$ on $N_c$, corresponding to the left and right y-axes, respectively. (b) Scaling plots of $|Z|/R^*$ versus $\omega/\omega^*$. Experimental results for 1 mm diameter spheres are shown in black dots with the fitted master curve marked with orange diamonds. Simulation results based on random packings with $N_c$ from 4 to 12 are presented by shaded error bars. Red dashed lines present typical master curves for 2D square lattice networks with $R$ and $C$ components. The AC response of simple $R//C$ equivalence is presented by the black dashed line. Each individual point shown in error bar is based on ten realizations.

From the comparison above, it can be concluded not only $P_c$ but also $N_c$ is important in determining the shape of the impedance vs frequency master curve of AC electrical transport in granular materials under various stress states. Conventional models generally comprise ordered 2D or 3D square lattices [22, 61, 62] with constant particle coordination, and furthermore do not incorporate the electrical conduction mechanisms intrinsic to granular materials. Consequently, such analyses tend not to reproduce the impedance-frequency scaling of granular systems with various packing structures and limit their utility towards the precise interpretation of observed transport phenomena. The present results demonstrate that by considering the topological configurations of conductive granular systems along with the physical characteristics of intergranular contacts, AC scaling behaviour in these systems can be meaningfully reproduced. The proposed framework sheds light on transport analyses of both electrons and ions in discrete materials and systems at various length scales, e.g., percolation-dominated conductivity [50, 55, 57, 59, 63], origins of NCL [4, 16, 64], dependence of AC electron/ion conductance curve on dimensionality [4, 5, 19, 30], etc.

## 4. Conclusion

We have presented here experimental observations of universal AC scaling behaviour in conductive granular systems under different stress states. Results reveal that impedance moduli curves collapse onto a unique master curve with $R^*$ and $\omega^*$ serving as scaling parameters, proving stress-independent universality. By validating our experimental results with the introduced numerical model based on 3D packing structures and equivalent RC networks, we have shown how both this scaling and the observed capacitive-to-resistive transition emerge as a consequence of percolation, which in turn is governed by network topology and inter-particle contact morphologies formed under a given stress level. Overall conduction in applied granular systems is thus fundamentally driven by the manner in which individual particles interact with each other in compression to give rise to resistive and capacitive elements and collectively produce percolation pathways. It flows from this that the mechanical and topological features of applied granular systems are key towards understanding their electrical responses.